\documentclass[12pt, twocolumn]{article}

\usepackage{natbib}
\usepackage{graphicx}
\usepackage{amsmath}
\usepackage{amsfonts}
\usepackage[
    a4paper,
    top=3.0cm,
    bottom=3.0cm,
    left=1.8cm,
    right=1.8cm
]{geometry}
\usepackage{enumitem}
\usepackage{xurl}      

\usepackage{nameref,hyperref}
\hypersetup{
   colorlinks=true,                   
   linkcolor=blue,                    
   urlcolor=blue,                     
   anchorcolor=blue,                  
   citecolor=blue,                    
}
\usepackage{tikz}
\usetikzlibrary{positioning,arrows.meta}

\begin{document}

   \onecolumn
	\section*{GravDyn: A Python Framework for Gravitational Modeling of Irregular Celestial Bodies Using Polyhedral, mascon, and Series Expansion Methods}

	Safwan Aljbaae\textsuperscript{1,*},
	valerio Carruba\textsuperscript{2, 3},
	Allan K. de Almeida Jr\textsuperscript{4},
	Gabriel Antonio Caritá\textsuperscript{5},
	Antonio F. B. A. Prado\textsuperscript{6},
	Marcelo L. Mota\textsuperscript{7}
	Carlos E. Ferreira Lopes\textsuperscript{7}\\
	\noindent
	{\scriptsize
		{\bf 1:} Universidad de Atacama, Instituto de Astronomía y Ciencias Planetarias, Copayapu 485, Copiapó, Chile.\\
		{\bf 2:} School of Natural Sciences and Engineering, São Paulo State University (UNESP), Guaratinguetá, SP 12516-410, Brazil\\
		{\bf 3:} Laboratório Interinstitucional de e-Astronomia, Rio de Janeiro, RJ 20765-000, Brazil\\
		{\bf 4:} CFisUC, Departamento de Física, Universidade de Coimbra, Coimbra, 3004-516, Portugal\\
		{\bf 5:} São Paulo State University (UNESP), Instituto de Geoci\^encias e Ci\^encias Exatas, Rio Claro, SP 13506-900, Brazil\\
		{\bf 6:} Divisão de Pós-Graduação, Instituto Nacional de Pesquisas Espaciais (INPE), Avenida dos Astronautas, 515, São José dos Campos, 12227-310, São Paulo, Brazil\\
		{\bf 7:} Federal Institute of São Paulo, IFSP, Avenida Thereza Ana Cecon Breda, s/n - Vila São Pedro, Hortol\^{a}ndia, 13183-250, SP, Brazil.\\
	}
	$^{*}$Corresponding author: \href{mailto:safwan.aljbaae@uda.cl}{safwan.aljbaae@uda.cl}
	\section*{Abstract}
		{\bf GravDyn} is an open‑source Python package for computing gravitational potentials and accelerations around irregular celestial bodies. It implements three approaches within the same workflow: the constant-density polyhedral method, a layered mascon approximation based on tetrahedral decomposition, and the Potential Series Expansion Method (PSEM). The package handles shape‑model preprocessing, builds the selected gravity representation, and evaluates potentials and accelerations efficiently. The three implemented methods serve different regimes: the polyhedral model provides a reference solution near the surface, the mascon model supports layered internal density structures, and PSEM offers fast evaluation outside the Brillouin sphere once the polynomial coefficients have been generated. Validation tests against the polyhedral solution show that the mascon and PSEM models reproduce the reference potential with small relative errors while reducing evaluation costs. {\bf GravDyn} is intended for mission design and studies of spacecraft motion, orbital stability, and gravitational modeling around asteroids and other small bodies.
		\section*{Keywords}
		celestial mechanics; asteroid dynamics; gravitational modeling; polyhedral gravity; mascon method; potential series expansion; celestial mechanics software; small-body environments

	\twocolumn

	\section{Overview}

	\subsection{Introduction}

		Irregularly shaped small bodies do not settle into hydrostatic equilibrium \citep{Dermott_1979}, so their gravity fields differ markedly from the simple point‑mass picture. Those differences matter when we design spacecraft trajectories, assess orbital stability, or plan missions that will operate close to an asteroid \citep{Scheeres_2012, Hao_2020, Ferrari_2021, Bottiglieri_2023}. Several approaches have been developed to address this problem. The classical polyhedral method provides an accurate analytical solution for the gravitational potential of a homogeneous body by transforming volume integrals into surface and line integrals, making it suitable for modeling the gravitational field near the surface of irregular bodies \citep{Werner_1997, Tsoulis_2001}. Despite its accuracy, this method is computationally expensive, especially for high-resolution shape models, as the number of faces and edges directly impacts the evaluation cost. Alternatively, a different philosophy replaces the continuous mass with a collection of point masses distributed throughout the volume \citep{Geissler_1996}. Building on that, later research formalized mascons using tetrahedral decompositions of a polyhedron. By placing a point mass at each tetrahedron’s centroid, this approach preserves much of the original density field while reducing the computational load \citep{Chanut_2015, Aljbaae_2017, Aljbaae_2021}. In addition, the method naturally allows the incorporation of layered internal structures with varying densities. More recently, the Potential Series Expansion Method (PSEM) has been proposed \citep{Mota_2023}. Here, the potential is expressed as a convergent polynomial series derived analytically from the tetrahedral elements. PSEM derives a polynomial expansion that provides a fast-to-evaluate approximation outside the Brillouin sphere.\\

		Although these formulations are well established, their implementations are often tied to individual studies. For instance, several tools for the polyhedral gravity model are available in different languages such as C++17 with Python interface \footnote{\href{https://github.com/esa/polyhedral-gravity-model}{https://github.com/esa/polyhedral-gravity-model}} or Fortran\footnote{\href{https://github.com/a-amarante/minor-gravity/tree/v2.1}{https://github.com/a-amarante/minor-gravity/tree/v2.1}}. This isolation between different methods makes it difficult to reproduce previous calculations, compare methods under the same assumptions, or reuse common shape-processing steps across projects. GravDyn addresses this practical gap by placing polyhedral, mascon, and PSEM gravity models in a single Python package. Each method presents different advantages and limitations depending on the application. {\bf GravDyn} helps users determine which method is most appropriate for a given application.\\

		The package follows a common workflow. A user provides a triangular shape model and the physical parameters of the body. {\bf GravDyn} then verifies and rescales the geometry, builds the selected gravity representation, and evaluates the potential or acceleration at prescribed field points. This paper describes the structure of the package, the implemented models, and the validation tests used to compare them.

	\subsection{Implementation and architecture}

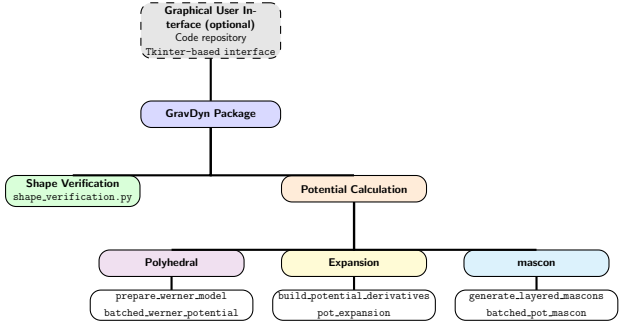
\begin{figure}[htbp]
		\centering
		\begin{tikzpicture}[
			scale=0.35, transform shape,
			font=\sffamily,
			>=Latex,
			box/.style={
				draw,
				rounded corners,
				align=center,
				minimum height=1cm,
				text width=5.2cm
			},
			smallbox/.style={
				draw,
				rounded corners,
				align=left,
				text width=6.0cm,
				inner sep=2pt
			},
			line/.style={-, thick}
		]

		\node[box, fill=blue!15, text width=5cm] (root) {\textbf{GravDyn Package}};

		\node[box, fill=gray!20, dashed, text width=5cm, above=1.6cm of root] (gui)
		{\textbf{Graphical User Interface (optional)}\\Code repository
		\texttt{Tkinter-based interface}};

		\draw[line] (gui.south) -- (root.north);

		\node[box, fill=green!15, below left=1.8cm and 0.02cm of root, text width=4.8cm] (shape)
		{\textbf{Shape Verification}\\ \texttt{shape\_verification.py}};

		\node[box, fill=orange!15, below right=1.8cm and 0.02cm of root, text width=5.2cm] (potcalc)
		{\textbf{Potential Calculation}};

		\draw[line] (root.south) |- (shape.north);
		\draw[line] (root.south) |- (potcalc.north);

		\node[box, fill=violet!12, below left=1.8cm and 1.4cm of potcalc] (poly)
		{\textbf{Polyhedral}};

		\node[box, fill=yellow!20, below=1.8cm of potcalc] (exp)
		{\textbf{Expansion}};

		\node[box, fill=cyan!12, below right=1.8cm and 1.4cm of potcalc] (mas)
		{\textbf{mascon}};

		\draw[line] (potcalc.south) |- (poly.north);
		\draw[line] (potcalc.south) -- (exp.north);
		\draw[line] (potcalc.south) |- (mas.north);

		\node[smallbox, below=0.5cm of poly, align=center] (polyf)
		{\texttt{prepare\_werner\_model}\\[4pt]
		\texttt{batched\_werner\_potential}};

		\node[smallbox, below=0.5cm of exp, align=center] (expf)
		{\texttt{build\_potential\_derivatives}\\[4pt]
		\texttt{pot\_expansion}};

		\node[smallbox, below=0.5cm of mas, align=center] (masf)
		{\texttt{generate\_layered\_mascons}\\[4pt]
		\texttt{batched\_pot\_mascon}};

		\draw[line] (poly.south) -- (polyf.north);
		\draw[line] (exp.south) -- (expf.north);
		\draw[line] (mas.south) -- (masf.north);

		\end{tikzpicture}
		\caption{Structure of the {\bf GravDyn} package. The software is organized into two main modules: shape verification and potential calculation. The potential calculation module includes polyhedral, expansion, and mascon-based approaches.}
		\label{fig:architecture}
		\end{figure}

		{\bf GravDyn} is implemented in Python and is organized into modular components. The package provides tools for efficient numerical evaluation of gravitational potentials and accelerations in small-body environments. As shown in Figure~\ref{fig:architecture}, the system consists of two pillars: a core computational library and an optional graphical user interface (GUI). The core lives in the \texttt{gravdyn} package and contains all numerical routines, data structures, and processing tools. The GUI offers a user-friendly interface for visualization, parameter configuration and interaction with the underlying models. The GUI is intentionally decoupled from the core computational layer. This separation ensures that the scientific functionality remains independent of the interface, allowing the software to be used both programmatically and interactively.

		\subsubsection{Shape-model preprocessing}

			The implementation is essentially based on a functional decomposition which constitutes the standard workflow for gravitational modeling. The first step is geometry preprocessing, which is handled by the \texttt{shape\_verification} module. This makes sure that the polyhedral shape model is geometrically and physically consistent before any gravitational analysis is performed. First the mesh is loaded and the body is aligned to its principal axes of inertia which is helpful to ensure that the following calculations are consistent with the physical structure of the object. The shape is then rescaled to match some reference volume derived from the specified mass and density and translated so that its center of mass coincides with the origin. Although these steps are standard from a scientific point of view, they are essential, as small inconsistencies at this stage can propagate into the dynamical results. The processed geometry is stored for later use, and a set of diagnostic plots is generated to verify that the preprocessing stage has been performed correctly.\\

			The second stage of the workflow is devoted to building the gravitational model and evaluating the resulting field. In {\bf GravDyn}, the polyhedral method, the mascon approach, and the PSEM are implemented within a single computational framework with a common interface. This design allows users to switch between gravity models, compare their performance and accuracy under the same conditions, and integrate different modeling strategies into the same dynamical workflow, rather than treating each method as a separate standalone implementation.\\

		\subsubsection{Polyhedral gravity model}

			The classical polyhedral method gives an exact analytical expression for the gravitational potential of a homogeneous body. In {\bf GravDyn}, this is done by combining the preprocessing routine \texttt{masconprepare\_werner\_model} with the evaluation function  \texttt{batched\_werner\_potential}. The preprocessing step follows the work of \citet{Werner_1997, Scheeres_2012}, to perform calculations characterizing the body, such as face and edge dyads, normal vectors, and other geometric coefficients. The polyhedral formulation generally provides the most accurate results, especially near the surface where the approximations are often not valid.\\

\begin{figure}[htbp]
				\centering
				\begin{tikzpicture}[scale=0.5, line join=round, line cap=round, >=Latex]

					\definecolor{facetorange}{RGB}{241,170,120}
					\definecolor{facetyellow}{RGB}{241,232,146}
					\definecolor{raygray}{RGB}{120,120,120}

					\coordinate (v1) at (-5.2,1.2);
					\coordinate (v2) at (-2.1,0.5);
					\coordinate (v3) at (-2.8,3.6);

					\coordinate (a1) at (-0.8,2.8);
					\coordinate (a2) at (-1.5,1.6);
					\coordinate (a3) at (0.1,1.9);

					\coordinate (y1) at (-1.45,1.55);
					\coordinate (y2) at (-0.25,1.90);
					\coordinate (y3) at (-0.95,2.55);

					\coordinate (O) at (5.2,-2.0);
					\coordinate (X) at (7.2,-2.0);
					\coordinate (Y) at (5.2,0.3);
					\coordinate (Z) at (3.9,-3.0);

					\draw[gray!70, thin] (O) -- (v1);
					\draw[gray!70, thin] (O) -- (v2);
					\draw[gray!70, thin] (O) -- (v3);

					\filldraw[fill=facetorange, draw=black, line width=1.2pt] (v1) -- (v2) -- (v3) -- cycle;

					\fill (v1) circle (2.6pt);
					\fill (v2) circle (2.6pt);
					\fill (v3) circle (2.6pt);

					\node[below left=2pt] at (v1) {$v_1$};
					\node[below right=2pt] at (v2) {$v_2$};
					\node[above=2pt] at (v3) {$v_3$};

					\node[] at (-3.5, 0.3)  {$E_{12}$};
					\node[] at (-1.9,2.0)  {$E_{23}$};
					\node[] at (-3.9,3.0)  {$E_{31}$};

					\draw[->, very thick] (-3.9,2.0) -- (-5.2,3.0);
					\node[above left] at (-5.2,3.0) {$\hat{\mathbf{n}}_{123}$};

					\fill (O) circle (2.8pt);

					\draw[->, very thick] (O) -- (X);
					\draw[->, very thick] (O) -- (Y);
					\draw[->, very thick] (O) -- (Z);

					\node[right] at (X) {$x$};
					\node[above] at (Y) {$y$};
					\node[below left] at (Z) {$z$};

				\end{tikzpicture}
				\caption{Redrawn schematic of the geometric entities associated with a triangular facet in the polyhedral gravity model of \citet{Werner_1997, Scheeres_2012}. The facet is defined by the ordered vertices $\mathbf{v}_1$, $\mathbf{v}_2$, and $\mathbf{v}_3$, with edges $E_{12}$, $E_{23}$, and $E_{31}$, and outward unit normal $\hat{\mathbf{n}}_{123}$. The field point is represented in the body-fixed frame $(x,y,z)$.}\label{fig:werner_polyhedron_geometry}
			\end{figure}
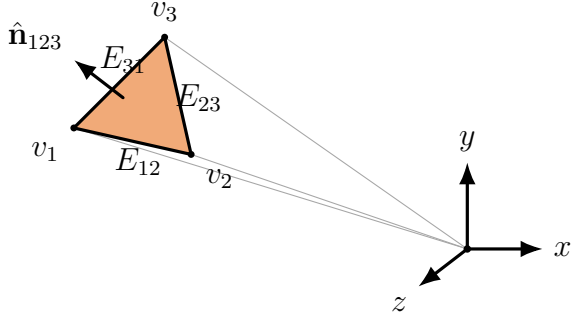

			The package implements the constant-density polyhedron formulation of \citet{Werner_1997, Scheeres_2012}. In this model, the body is described by a closed triangular mesh, and the gravitational potential is evaluated as a sum of edge and face contributions. For a field point $\mathbf{r}$, we define $\mathbf{r}_{e}$ as the vector from an arbitrary point on edge $e$ to $\mathbf{r}$, and $\mathbf{r}_{f}$ as the vector from an arbitrary point on face $f$ to $\mathbf{r}$. Each triangular facet is specified by three ordered vertices, for example $\mathbf{v}_1$, $\mathbf{v}_2$, and $\mathbf{v}_3$, with associated edges $E_{12}$, $E_{23}$, and $E_{31}$, and an outward unit normal $\hat{\mathbf{n}}_{123}$. A schematic representation of these geometric quantities is shown in Fig.~\ref{fig:werner_polyhedron_geometry}.

			The potential can then be written as
			\begin{equation}
				\resizebox{0.8\linewidth}{!}{$
				\begin{aligned}
				U(\mathbf{r}) =
				\frac{G\sigma}{2}
				\left[
				\sum_{e \in \mathrm{edges}}
				\mathbf{r}_{e}\mathbf{E}_{e}\mathbf{r}_{e} L_{e}
				-
				\sum_{f \in \mathrm{faces}}
				\mathbf{r}_{f}\mathbf{F}_{f}\mathbf{r}_{f} \omega_{f}
				\right]
				\end{aligned}
				$}
			\end{equation}

			The corresponding first and second derivatives are

			\begin{eqnarray}
				\resizebox{0.8\linewidth}{!}{$
				\begin{aligned}
				\nabla U(\mathbf{r}) &=&
				-G\sigma
				\left[
				\sum_{e \in \mathrm{edges}}
				\mathbf{E}_{e}\mathbf{r}_{e} L_{e}
				-
				\sum_{f \in \mathrm{faces}}
				\mathbf{F}_{f}\mathbf{r}_{f} \omega_{f}
				\right],\\
				\nabla^{2}_{\mathbf{r}} U(\mathbf{r}) &=&
				G\sigma
				\left[
				\sum_{e \in \mathrm{edges}}
				\mathbf{E}_{e} L_{e}
				-
				\sum_{f \in \mathrm{faces}}
				\mathbf{F}_{f} \omega_{f}
				\right].
				\end{aligned}
				$}
			\end{eqnarray}

			where $G$ is the gravitational constant and $\sigma$ is the bulk density. The matrices $\mathbf{E}_{e}$ and $\mathbf{F}_{f}$ are the edge and face dyads, respectively, while $L_{e}$ is the logarithmic edge factor and $\omega_{f}$ is the signed solid angle subtended by face $f$ at the evaluation point. These quantities are defined as
			\begin{eqnarray}
				\resizebox{0.8\linewidth}{!}{$
				\begin{aligned}
				\resizebox{\linewidth}{!}{$
				\begin{aligned}
				\mathbf{E}_{e}
				&=&
				\hat{\mathbf{n}}_{f}\hat{\mathbf{n}}_{e}^{f}
				+
				\hat{\mathbf{n}}_{f'}\hat{\mathbf{n}}_{e}^{f'},\\
				\mathbf{F}_{f}
				&=&
				\hat{\mathbf{n}}_{f}\hat{\mathbf{n}}_{f},\\
				L_{e}
				&=&
				\ln
				\left(
				\frac{r_{1}^{e}+r_{2}^{e}+e_{e}}
					{r_{1}^{e}+r_{2}^{e}-e_{e}}
				\right),\\
				\omega_{f}
				&=&
				2 \arctan
				\left(
				\frac{
				\mathbf{r}_{1}^{f}\cdot
				\mathbf{r}_{2}^{f}\cdot\mathbf{r}_{3}^{f}
				}{
				r_{1}^{f}r_{2}^{f}r_{3}^{f}
				+
				r_{1}^{f}\mathbf{r}_{2}^{f}\cdot\mathbf{r}_{3}^{f}
				+
				r_{2}^{f}\mathbf{r}_{3}^{f}\cdot\mathbf{r}_{1}^{f}
				+
				r_{3}^{f}\mathbf{r}_{1}^{f}\cdot\mathbf{r}_{2}^{f}
				}
				\right).
				\end{aligned}
				$}
				\end{aligned}
				$}
			\end{eqnarray}

			Here $f$ and $f'$ are the two faces adjacent to edge $e$, $\hat{\mathbf{n}}_{f}$ and $\hat{\mathbf{n}}_{f'}$ are their outward unit normals, and $\hat{\mathbf{n}}_{e}^{f}$ and $\hat{\mathbf{n}}_{e}^{f'}$ are the corresponding edge-normal unit vectors lying in each face and pointing outward from the face. The scalar $e_{e}$ is the length of edge $e$, while $r_{1}^{e}$ and $r_{2}^{e}$ are the distances from the evaluation point to the two vertices of that edge. Similarly, $\mathbf{r}_{1}^{f}$, $\mathbf{r}_{2}^{f}$, and $\mathbf{r}_{3}^{f}$ are the vectors from the evaluation point to the three vertices of face $f$, with $r_{i}^{f}=\|\mathbf{r}_{i}^{f}\|$.

			In the preprocessing stage, {\bf GravDyn} therefore computes and stores the geometric quantities required by these sums: face normals, edge normals, face dyads, edge dyads, edge lengths, and the connectivity between adjacent faces and edges. During field evaluation, these precomputed quantities are reused so that only the point-dependent terms $L_{e}$, $\omega_{f}$, $\mathbf{r}_{e}$, and $\mathbf{r}_{f}$ must be evaluated. This is also why the cost of the polyhedral method scales with the number of faces and edges in the mesh. The formulation is especially valuable near the surface because it remains valid down to the boundary of the body, unlike spherical-harmonic expansions, which may diverge close to or inside the circumscribing sphere.\\

			The numerical evaluation of closed-form polyhedral gravity models near singular geometrical configurations was analyzed by \citet{Tsoulis_2001}. Their work complements the formulation of \citet{Werner_1997} by treating the cases in which the orthogonal projection of the computation point onto a face lies inside the polygon, on an edge, or at a vertex. In such situations, apparent singularities arise in the line-integral expressions, although the potential and first-order derivatives remain well defined when appropriate limiting correction terms are included. Incorporating this treatment would require additional geometrical tests and correction terms during the field evaluation, increasing the computational cost of the method. The present version of {\bf GravDyn} therefore keeps the \citet{Werner_1997} edge-face formulation as the baseline polyhedral model, while the explicit \citet{Tsoulis_2001} singularity treatment is left for a future implementation stage.\\

		\subsubsection{Layered mascon model}

			The mascon gravity model with a polyhedral source is implemented by the functions \texttt{generate\_layered\_mascons} and \texttt{batched\_pot\_mascon}. As in \citet{Chanut_2015,Aljbaae_2017,Aljbaae_2021}, we connect each face of the surface mesh to the body-fixed origin to form a tetrahedron. If the three vertices of a face are denoted by $\mathbf{v}_1$, $\mathbf{v}_2$, and $\mathbf{v}_3$, then the signed volume of the associated tetrahedron is obtained as
			\begin{equation}
				V_T =
				\frac{1}{6}
				\mathbf{v}_1 \cdot
				\left(\mathbf{v}_2 \times \mathbf{v}_3\right).
			\end{equation}
			The tetrahedron is then subdivided radially according to the number of density layers provided by the user. For a model with $N_{\ell}$ layers, the scaled vertices of the $k$-th radial boundary are written as
			\begin{equation}
				\mathbf{v}_{j,k}
				=
				\lambda_k \mathbf{v}_j,
				\qquad
				\lambda_k = \frac{k}{N_{\ell}},
				\qquad
				j=1,2,3.
			\end{equation}
			The cumulative volume enclosed by this scaled tetrahedron is
			\begin{equation}
				V_{T,k}
				=
				\frac{1}{6}
				\mathbf{v}_{1,k}\cdot
				\left(\mathbf{v}_{2,k}\times\mathbf{v}_{3,k}\right),
			\end{equation}
			so that the volume assigned to layer $k$ is
			\begin{equation}
				\Delta V_{T,k}
				=
				\begin{cases}
				V_{T,1}, & \text{if } k=1,\\
				V_{T,k}-V_{T,k-1}, & \text{if } k>1.
				\end{cases}
			\end{equation}
			The preliminary mass and gravitational parameter of each mascon are then computed from the layer density as
			\begin{equation}
				m_{T,k}^{0} = \rho_k \Delta V_{T,k},
				\qquad
				\mu_{T,k}^{0} = G m_{T,k}^{0}.
			\end{equation}

			The mascon position is assigned using the same approximation adopted in the original layered implementation. For the innermost layer, the point mass is placed at
			\begin{equation}
				\mathbf{c}_{T,1}
				=
				\frac{
				\mathbf{v}_{1,1}+\mathbf{v}_{2,1}+\mathbf{v}_{3,1}
				}{4},
			\end{equation}
			while for the outer layers it is placed at the average position of the three
			vertices of the outer and inner layer boundaries,
			\begin{equation}
				\resizebox{0.85\linewidth}{!}{$
				\begin{aligned}
				\mathbf{c}_{T,k}
				=
				\frac{
				\mathbf{v}_{1,k}+\mathbf{v}_{2,k}+\mathbf{v}_{3,k}
				+
				\mathbf{v}_{1,k-1}+\mathbf{v}_{2,k-1}+\mathbf{v}_{3,k-1}
				}{6},
				\qquad k>1.
				\end{aligned}
				$}
			\end{equation}
			After all mascons have been generated, the total preliminary mass is compared with the prescribed asteroid mass $M$. If necessary, all masses are rescaled by a common factor,
			\begin{equation}
				\resizebox{0.85\linewidth}{!}{$
				\begin{aligned}
				\alpha =
				\frac{M}{\sum_i m_i^{0}},
				\qquad
				m_i = \alpha m_i^{0},
				\qquad
				\mu_i = Gm_i.
				\end{aligned}
				$}
			\end{equation}
			This step conserves the total mass of the body,
			\begin{equation}
				\sum_i m_i = M,
			\end{equation}
			while preserving the relative mass distribution imposed by the selected density
			layers. The potential at a field point $\mathbf{r}$ is then approximated as
			\begin{equation}
				U(\mathbf{r})
				=
				\sum_{i=1}^{N_m}
				\frac{\mu_i}{r_i},
				\qquad
				r_i =
				\left\|
				\mathbf{r}-\mathbf{c}_i
				\right\|,
			\end{equation}
			where $N_m$ is the total number of mascons. The function \texttt{batched\_pot\_mascon} evaluates these point-mass contributions for many field points simultaneously, making the method suitable for large grids and long orbital integrations. The main advantage of this implementation is that it keeps the geometry tied to the original polyhedral shape, supports layered internal density models, and preserves the prescribed total mass.

		\subsubsection{Potential Series Expansion Method}

			The Potential Series Expansion Method (PSEM) is based on the analytical expansion of the Newtonian potential kernel $1/r$ into a convergent power series of Legendre polynomials. Considering a field point $\mathbf{r}$ and a source point inside the body, the inverse distance can be written as
			\begin{equation}
				\frac{1}{r}
				=
				\frac{1}{\rho}
				\sum_{i=0}^{\infty}
				P_i(u)
				\left(\frac{\rho'}{\rho}\right)^i,
			\end{equation}
			where $\rho=\|\mathbf{r}\|$, $\rho'$ is the distance from the origin to the mass element, and $u=\cos\gamma$, with $\gamma$ being the angle between the vectors $\mathbf{r}$ (from the origin to the field point) and $\mathbf{r}'$ (from the origin to the mass element). This angle satisfies
			\begin{equation}
			\cos\gamma =
			\frac{\mathbf{r}\cdot\mathbf{r}'}{\|\mathbf{r}\|\,\|\mathbf{r}'\|}.
			\end{equation}
			This expansion is uniformly convergent for points located outside the Brillouin sphere of the body.

			In PSEM, the irregular body is first represented as a polyhedron and decomposed into tetrahedral elements, similarly to the classical polyhedral method. The total gravitational potential is then obtained by integrating the series term-by-term over each tetrahedron and summing all contributions, yielding
			\begin{equation}
			U(\mathbf{r})
			=
			G\sigma
			\sum_{k=1}^{N_T}
			\sum_{i=0}^{m}
			\iiint_{Q_k}
			P_i(u)
			\frac{\rho'^i}{\rho^{i+1}}
			\, dV,
			\end{equation}
			where $Q_k$ denotes each tetrahedral element, $N_T$ is the total number of tetrahedra, and $m$ is the truncation order of the expansion. As shown in \citet{Mota_2023}, each term of the integrand becomes a polynomial function of the coordinates, allowing the volume integrals to be evaluated analytically using a coordinate transformation to a reference tetrahedron. The resulting potential can be expressed as a finite polynomial expansion in Cartesian coordinates of the form
			\begin{equation}\label{eq:PSEM}
				\resizebox{0.85\linewidth}{!}{$
				\begin{aligned}
			U(\mathbf{r})
			=
			\sum_{l=0}^{m}
			\sum_{i+j+k=l}
			U^{l}_{i,j,k}
			\,
			x^i y^j z^k
			\, r^{-(2l+1)/2} ~ \\
			\forall~i,j,k,l \in \mathbb{N}:i+j+k=l,
				\end{aligned}
				$}
			\end{equation}
			which enables fast evaluation of the potential and its derivatives.\\

			In {\bf GravDyn}, the function \texttt{build\_potential\_derivatives} is used to obtain the PSEM coefficients. Deriving these coefficients symbolically becomes computationally expensive for high truncation orders and high-resolution shape models. The computational cost depends strongly on the number of faces in the polyhedral shape model, the corresponding number of tetrahedral elements, and the selected truncation order. For some high-resolution asteroid models, the coefficient-generation stage may require several days of computation.\\

			For this reason, the current release of {\bf GravDyn} provides PSEM as a runtime evaluation tool based on precomputed and validated coefficient files. These polynomial expansions are generated offline on dedicated servers and stored in the {\bf GravDyn} repository for selected bodies. At runtime, \texttt{build\_potential\_derivatives} loads the precomputed expansion for the selected asteroid and returns both the symbolic expressions and optimized numerical callables, enabling very fast evaluation of the potential and its derivatives from Eq.~\eqref{eq:PSEM}.\\

			If a precomputed model is not yet available, the function notifies the user and provides instructions to request its generation. The generation of new coefficient sets is currently maintained as an offline preprocessing workflow by the authors, rather than as a standard user-facing routine, because each new model requires substantial symbolic computation and validation against reference solutions. Once generated and validated, the corresponding model can be added to the repository for future reuse by the community.\\

	\subsection{Quality control}

		\subsubsection{Automated tests}

			{\bf GravDyn} includes automated tests written with \texttt{pytest}. These cover shape preprocessing, polyhedral precomputation, mascon generation, PSEM model loading, plotting utilities, and point-mass evaluations against analytical solutions. The aim of the test suite is to verify that the implemented routines return consistent arrays, reuse cached data correctly, and reproduce reference values for representative cases.\\

		\subsubsection{Input checks}

			Input validation is applied before field evaluation. The package checks array dimensions, evaluation-point shapes, mass and density inputs, and the availability of external data files, and it raises informative errors when a configuration is inconsistent.

		\subsubsection{Validation and performance}

			{\bf GravDyn} is validated through unit-level tests and also by comparing the gravitational modeling approaches against the classical polyhedral solution, which is taken as the reference. The mascon and PSEM implementations are evaluated at varying distances from the body.\\

\begin{figure}[htbp]
				\centering
				\includegraphics[width=0.88\linewidth]{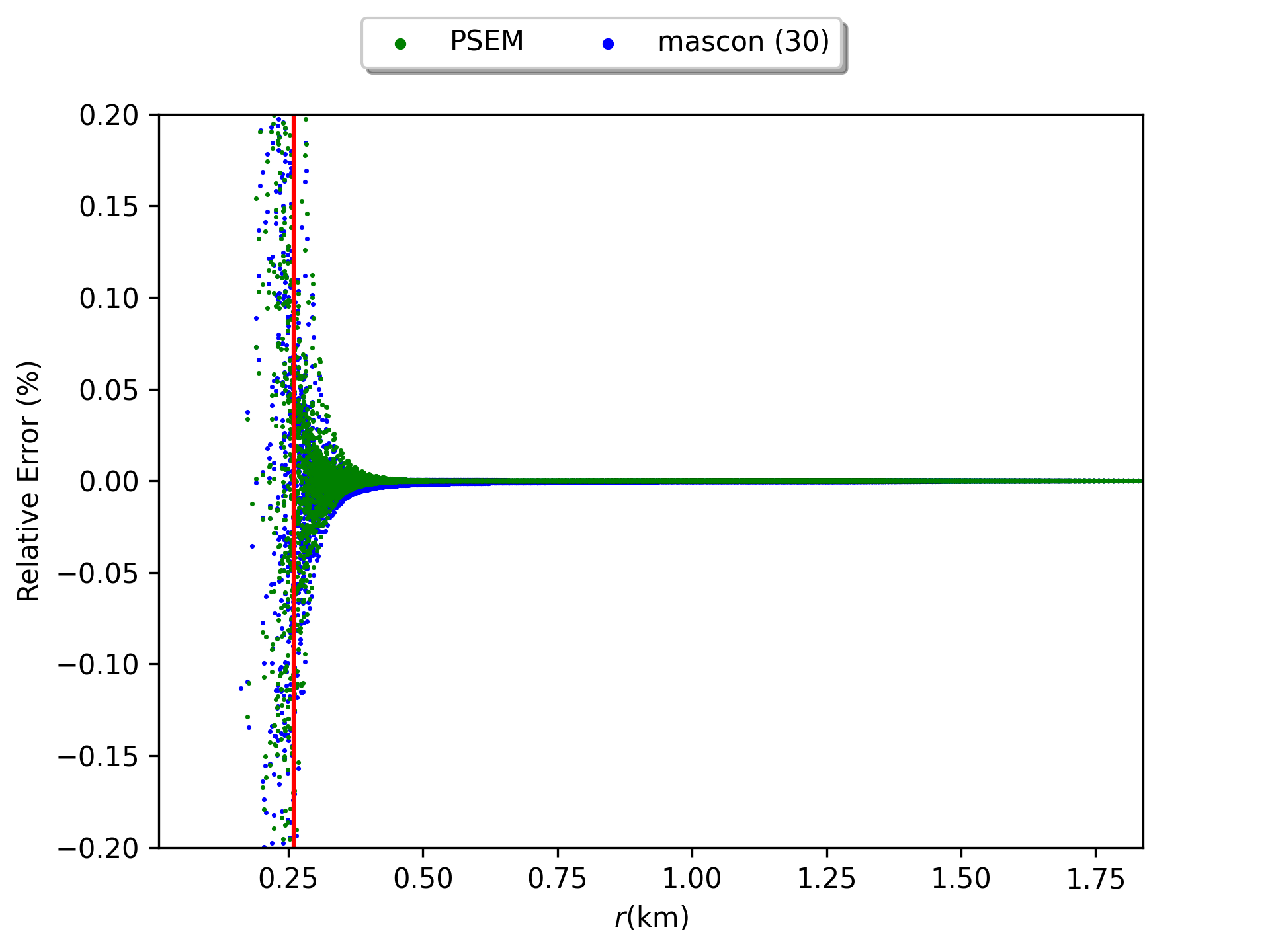}
				\caption{Example comparison of relative error in the gravitational potential as a function of radial distance $r$. The polyhedral method is used as a reference solution. The expansion method (green) and mascon approach (blue) show convergence toward the reference as the distance increases. The vertical red line indicates the approximate radius of the body.}
				\label{fig:method_comparison_example}
			\end{figure}

			Figure~\ref{fig:method_comparison_example} compares the mascon and PSEM models against the polyhedral reference as a function of radial distance. Both approximations converge toward the reference in the external region, while Table~\ref{tab:performance_comparison} shows the expected cost trade-off: PSEM is fastest once its coefficients have been precomputed, whereas the mascon model remains useful when layered interiors must be represented explicitly. The results shown in Figure~\ref{fig:method_comparison_example} and Table~\ref{tab:performance_comparison} refer specifically to asteroid (99942) Apophis, which is used here as an illustrative benchmark. Similar comparisons for other bodies, including (21) Lutetia and (87) Sylvia, have been performed in previous work \citep{Mota_2023}. The row labelled ``Tsoulis'' is included only as an external reference benchmark. It does not correspond to a released \texttt{gravdyn} module, but to a standalone Fortran implementation of the polyhedral gravity formulation following Tsoulis and Petrovi{\'c} \citep{Tsoulis_2001}. This implementation was not written using the vectorized, parallel, or JAX-based execution paths used in the Python routines. It is therefore not intended as a direct software-performance comparison with the released package, but as a reference point for a well-established and robust polyhedral formulation. \\

			\begin{table}[htbp]
				\centering

				\caption{Performance comparison of gravitational modeling methods for asteroid (99942) Apophis. The test consists of evaluating the gravitational potential at 40{,}000 field points using a shape model with 3{,}996 faces. Timings were obtained on a system equipped with an Intel(R) Core(TM) i7-14700 processor (20 physical cores, 28 logical CPUs), with a maximum clock frequency of 5.4~GHz and 33~MiB of L3 cache.}

				\label{tab:performance_comparison}
				\resizebox{0.50\textwidth}{!}{
					\begin{tabular}{lccc}
						\hline
						\textbf{\scriptsize Method} & \textbf{\scriptsize Execution Time (s)} & \textbf{\scriptsize Relative Speed} & \textbf{\scriptsize Mean Relative Error (\%)} \\
						\hline
						Tsoulis        & 155.21 & $\times 1$   & Reference \\
						Werner         & 9.96   & $\times 15.6$ & Reference \\
						Mascon (30) & 20.14  & $\times 7.7$  & $6.86\times10^{-5}$ \\
						PSEM           & 1.50   & $\times 103.5$ & $1.84\times10^{-4}$\\
						\hline
					\end{tabular}
				}
			\end{table}
			The ``Mean Relative Error'' reported in Table~\ref{tab:performance_comparison} is the average of the pointwise relative potential errors shown as a function of radial distance in Figure~\ref{fig:method_comparison_example}, computed with respect to the polyhedral reference solution over the 40,000 evaluation points.\\

			The PSEM evaluation is essentially the direct computation of an analytical polynomial, which makes it very fast and inexpensive to evaluate, even when higher orders are required. The mascon approach, by contrast, involves summing the contributions of a large number of discrete masses; this becomes increasingly demanding as the number of layers grows, particularly when a refined internal structure is modeled. Very close to the surface, the mascon formulation can behave more steadily provided the discretization is sufficiently fine, whereas PSEM becomes less reliable due to truncation effects and its formal validity being restricted to the region outside the Brillouin sphere\footnote{The Brillouin sphere is defined as the smallest sphere centered at the center of mass that completely encloses the body. Series expansions of the gravitational potential, such as spherical harmonics or PSEM, are guaranteed to converge only outside this sphere; inside it, the series may diverge or lose accuracy (see, e.g., \citet{Heiskanen_1967}).}. Even so, both methods reach a level of accuracy that is more than adequate from a dynamical perspective: the residual errors are small enough that they do not meaningfully affect the qualitative or quantitative evolution of orbital trajectories in practical short-term trajectory integrations.\\

		\subsubsection{Reproducibility and execution environment}

			For general use, no special hardware is required. For computationally intensive applications, in particular for generating mascon models with multiple layered internal structure, it is advantageous to have a system with sufficient memory (suggested $\geq$ 16 GB RAM) and multi-core CPU support. The generation of layered mascon distributions involves processing large polyhedral meshes and assigning mass elements for multiple density layers, which can greatly increase memory usage and computational time. The required disk space depends on the size of the polyhedral shape model and associated data sets. Typically, less than 1 GB of space is required, but larger shape models and pre-computed data sets may require more.\\

			For typical use, {\bf GravDyn} does not require specialized hardware. Larger layered-mascon calculations benefit from additional memory and multi-core CPUs. Detailed system requirements are summarized in the Availability section.

	\section{Availability}

		\vspace{0.5cm}

		\subsection{Operating system}

			{\bf GravDyn} is platform-independent and has been tested on Linux (Ubuntu 22.04 and later). The software is expected to run on macOS and Windows systems with a compatible Python environment.\\

		\subsection{Programming language}

			Python (version 3.9 or higher)\\

		\subsection{System requirements}

			For general use, no special hardware is required. For computationally intensive applications, in particular for generating mascon models with multiple layered internal structure, it is advantageous to have a system with sufficient memory (suggested $\geq$ 16 GB RAM) and multi-core CPU support. The generation of layered mascon distributions involves processing large polyhedral meshes and assigning mass elements for multiple density layers, which can greatly increase memory usage and computational time. Disk space requirements depend on the size of the polyhedral shape model and the associated data sets. Typical use requires less than 1 GB of space, but larger shape models and precomputed data sets may require more.\\

		\subsection{Dependencies}

			{\bf GravDyn} relies on the following core Python libraries:
			\begin{itemize}
				\item NumPy (numerical computations)
				\item SymPy (symbolic computation for PSEM)
				\item Matplotlib (visualization)
				\item Trimesh (geometry handling and mesh processing)
			\end{itemize}

			Optional dependencies:
			\begin{itemize}
				\item JAX (accelerated computation and automatic differentiation)
				\item Pandas (data handling for mascon datasets)
			\end{itemize}

			All dependencies can be installed via standard package managers such as pip.\\

		\subsection{List of contributors}

			\begin{itemize}

				\item {\bf Safwan Aljbaae}\\
				Conceived and led the project, developed the {\bf GravDyn} software, and wrote the manuscript.\\

				\item {\bf Valerio Carruba}\\
				Contributed to testing and validation of the software, assisted in refining the implementation, and contributed to the revision and improvement of the manuscript.\\

				\item {\bf Allan K. de Almeida Jr}\\
				Contributed to testing and validation of the software, assisted in refining the implementation, and contributed to the revision and improvement of the manuscript.\\

				\item {\bf Gabriel Caritá}\\
				Contributed to testing and validation of the software, assisted in refining the implementation, and contributed to the revision and improvement of the manuscript.\\

				\item {\bf Antonio F. B. A. Prado}\\
				Contributed to testing and validation of the software, assisted in refining the implementation, and contributed to the revision and improvement of the manuscript.\\

				\item {\bf Marcelo, L. Mota}\\
				Contributed to the theoretical development of the Potential Series Expansion Method (PSEM).\\

				\item {\bf Carlos E. Ferreira Lopes}\\
				Contributed to testing and validation of the software, assisted in refining the implementation, and contributed to the revision and improvement of the manuscript.\\

			\end{itemize}

		\subsection{Software location:}

			{\bf Archive}

				\begin{description}[noitemsep,topsep=0pt]
					\item[Name:] Zenodo.
					\item[Persistent identifier:] \href{https://doi.org/10.5281/zenodo.20210316}{https://doi.org/10.5281/zenodo.20210316}.
					\item[Licence:] MIT License.
					\item[Publisher:] Zenodo.
					\item[Version published:] v0.1.2.
					\item[Date published:] 15-07-2026.
				\end{description}

			{\bf Code repository}

				\begin{description}[noitemsep,topsep=0pt]
					\item[Name:] {\bf GravDyn}.
					\item[Identifier:] {\scriptsize \href{https://github.com/safwanaljbaae/GravDyn.git}{https://github.com/safwanaljbaae/GravDyn.git}}
					\item[Licence:] MIT License.
					\item[Date published:] 2026-04-20.
				\end{description}

			{\bf Project website}

				The official {\bf GravDyn} website is available at \url{https://gravdyn.linea.org.br}. The website is hosted by the
				Laboratório Interinstitucional de e-Astronomia (LIneA).

			{\bf Package distribution}

				{\bf GravDyn} is also distributed via the Python Package Index (PyPI) and can be installed using:

				\begin{verbatim}
				pip install gravdyn
				\end{verbatim}

		\subsection{Language}

			All documentation is provided in English.\\

	\section{Reuse potential}

		{\bf GravDyn} can be reused in studies that require gravitational fields for irregular small bodies. The most direct applications are spacecraft trajectory design, orbital stability analysis, equilibrium-point calculations, and close-proximity motion near asteroids. Because the same input shape can be used to build polyhedral, mascon, and PSEM models, the package is also useful for method comparison under common geometric and physical assumptions.\\

		The layered mascon implementation is particularly useful when internal structure matters. By assigning densities to radial layers, users can test how simplified heterogeneous interiors modify the potential and acceleration field.\\

		The PSEM implementation serves a different role. Once its coefficients have been generated and stored, the potential and its derivatives can be evaluated rapidly outside the Brillouin sphere. This makes the method useful for large grids of field points and repeated evaluations in long integrations.\\

		{\bf GravDyn} can also serve as a validation environment for new approximations, since the polyhedral model is available as a reference solution within the same workflow. Additional precomputed PSEM models may be added to the repository as they are generated and validated.\\

		Users can report bugs, request support, or suggest improvements through the GitHub issue tracker associated with the GravDyn repository. For scientific questions or collaboration requests, users may also contact the corresponding author.\\

	\section*{Acknowledgements}

		The authors acknowledge the use of AI-based writing tools to assist with language editing and readability. All scientific content, results, and interpretations remain the responsibility of the authors. The official GravDyn project website (\url{https://gravdyn.linea.org.br}) is hosted by the Laboratório Interinstitucional de e-Astronomia (LIneA), whose infrastructure is supported by FINEP, CNPq, FAPERJ, and the INCT do e-Universo program. GAC is grateful to the São Paulo Research Foundation (FAPESP), grant 2025/07596-3. AKAJ acknowledges support from project SPACE, ref. COMPETE2030-FEDER-00860300, funded by COMPETE 2030 and FCT, Portugal.

	\section*{Competing interests}

		The authors have no competing interests to declare.

\bibliographystyle{plainnat}
\bibliography{references}

@article{Scheeres_2012,
author = {Scheeres, D.},
year = {2012},
month = {01},
pages = {},
title = {Orbital Motion in Strongly Perturbed Environments},
isbn = {978-3-642-03255-4},
journal = {Orbital Motion in Strongly Perturbed Environments, by Scheeres, Daniel J. ISBN: 978-3-642-03255-4. Berlin: Springer, 2012},
doi = {10.1007/978-3-642-03256-1}
}

@article{Werner_1997,
author = {Werner, Robert and Scheeres, D.},
year = {1996},
month = {09},
pages = {313-344},
title = {Exterior Gravitation of a Polyhedron Derived and Compared with Harmonic and Mascon Gravitation Representations of Asteroid 4769 Castalia},
volume = {65},
journal = {Celestial Mechanics and Dynamical Astronomy},
doi = {10.1007/BF00053511}
}

@article{Tsoulis_2001,
author = {Tsoulis, Dimitrios and Petrovic, Sveto},
year = {2001},
month = {03},
pages = {},
title = {On the singularities of the gravity field of a homogeneous polyhedral body},
volume = {66},
journal = {Geophysics},
doi = {10.1190/1.1444944}
}

@article{Geissler_1996,
author = {Geissler, Paul and Petit, J.-M and Durda, Daniel and Greenberg, Richard and Bottke, William and Nolan, Michael and Moore, Jeffrey},
year = {1996},
month = {03},
pages = {140-157},
title = {Erosion and Ejecta Reaccretion on 243 Ida and Its Moon},
volume = {120},
journal = {Icarus},
doi = {10.1006/icar.1996.0042}
}

@ARTICLE{Chanut_2015,
   author = {{Chanut}, T.~G.~G. and Aljbaae,S. and {Carruba}, V.},
    title = "{Mascon gravitation model using a shaped polyhedral source}",
  journal = {MNRAS},
     year = 2015,
    month = jul,
   volume = 450,
    pages = {3742-3749},
      doi = {10.1093/mnras/stv845},
   adsurl = {http://adsabs.harvard.edu/abs/2015MNRAS.450.3742C}
}

@ARTICLE{Aljbaae_2017,
   author = {{Aljbaae}, S. and {Chanut}, T.~G.~G. and {Carruba}, V. and {Souchay}, J. and
	{Prado}, A.~F.~B.~A. and {Amarante}, A.},
    title = "{The dynamical environment of asteroid 21 Lutetia according to different internal models}",
  journal = {MNRAS},
archivePrefix = "arXiv",
   eprint = {1610.02338},
 primaryClass = "astro-ph.EP",
     year = 2017,
    month = jan,
   volume = 464,
    pages = {3552-3560},
      doi = {10.1093/mnras/stw2619},
   adsurl = {http://adsabs.harvard.edu/abs/2017MNRAS.464.3552A}
}

@ARTICLE{Aljbaae_2021,
       author = {{Aljbaae}, {S.} and {Souchay}, J. and {Carruba}, V. and {Sanchez}, D.~M. and {Prado}, A.~F.~B.~A.},
        title = "{Influence of Apophis' spin axis variations on a spacecraft during the 2029 close approach with Earth}",
      journal = {Romanian Astronomical Journal},
         year = 2021,
        month = nov,
       volume = {31},
       number = {3},
        pages = {317-337},
archivePrefix = {arXiv},
       eprint = {2105.14001},
 primaryClass = {astro-ph.EP},
       adsurl = {https://ui.adsabs.harvard.edu/abs/2021RoAJ...31..317A}
}

@ARTICLE{Mota_2023,
       author = {{Mota}, M.~L. and {Aljbaae}, S. and {Prado}, A.~F.~B.~A.},
        title = "{The potential series expansion method: application to the asteroid (87) Sylvia}",
      journal = {European Physical Journal Special Topics},
         year = 2023,
        month = dec,
       volume = {232},
       number = {18-19},
        pages = {2961-2966},
          doi = {10.1140/epjs/s11734-023-01026-w},
       adsurl = {https://ui.adsabs.harvard.edu/abs/2023EPJST.232.2961M}
}

@article{Dermott_1979,
title = {Shapes and gravitational moments of satellites and asteroids},
journal = {Icarus},
volume = {37},
number = {3},
pages = {575-586},
year = {1979},
issn = {0019-1035},
doi = {https://doi.org/10.1016/0019-1035(79)90015-0},
url = {https://www.sciencedirect.com/science/article/pii/0019103579900150},
author = {Stanley F. Dermott}
}

@ARTICLE{Ferrari_2021,
       author = {{Ferrari}, Fabio and {Franzese}, Vittorio and {Pugliatti}, Mattia and {Giordano}, Carmine and {Topputo}, Francesco},
        title = "{Trajectory Options for Hera's Milani CubeSat Around (65803) Didymos}",
      journal = {Journal of the Astronautical Sciences},
         year = 2021,
        month = dec,
       volume = {68},
       number = {4},
        pages = {973-994},
          doi = {10.1007/s40295-021-00282-z},
       adsurl = {https://ui.adsabs.harvard.edu/abs/2021JAnSc..68..973F}
}

@article{Bottiglieri_2023,
author = {Bottiglieri, Claudio and Piccolo, Felice and Giordano, Carmine and Ferrari, Fabio and Topputo, Francesco},
year = {2023},
month = {05},
pages = {464},
title = {Applied Trajectory Design for CubeSat Close-Proximity Operations around Asteroids: The Milani Case},
volume = {10},
journal = {Aerospace},
doi = {10.3390/aerospace10050464}
}

@article{Hao_2020,
  title={Orbital maneuver strategy design based on piecewise linear optimization for spacecraft soft landing on irregular asteroids},
  author={Zhiwei Hao and Yi Zhao and Ying Chen and Qiuhua Zhang},
  journal={Chinese Journal of Aeronautics},
  year={2020},
  volume={33},
  pages={2694-2706},
  url={https://api.semanticscholar.org/CorpusID:213327572}
}

@book{Heiskanen_1967,
  author = {Heiskanen, W. A. and Moritz, H.},
  title = {Physical Geodesy},
  publisher = {W. H. Freeman},
  year = {1967}
}

\vspace{2cm}

\end{document}